\documentclass[10pt,letterpaper]{article}
\usepackage[top=0.75in,left=0.75in,footskip=0.75in,marginparwidth=1in]{geometry}

\usepackage[utf8]{inputenc}

\raggedright
\usepackage{lastpage,fancyhdr,graphicx}
\usepackage{epstopdf}
\usepackage{amsmath}
\usepackage{amssymb}
\usepackage[aboveskip=1pt,labelfont=bf,labelsep=period,singlelinecheck=off]{caption} 
\usepackage{changepage} 
\usepackage{cite} 
\usepackage{float} 
\usepackage{graphicx} 
\usepackage{microtype} 
\DisableLigatures[f]{encoding = *, family = * }
\usepackage{nameref,hyperref} 
\usepackage{sidecap} 
\usepackage{siunitx} 
\usepackage{subfig} 
\usepackage{color} 
\usepackage[dvipsnames]{xcolor} 

\usepackage{wrapfig}
\usepackage[pscoord]{eso-pic}
\usepackage[fulladjust]{marginnote}
\reversemarginpar

\makeatletter
\renewcommand{\@biblabel}[1]{\quad#1.}
\makeatother

\newcommand{\Pen}[0]{\textnormal{Pe}}

\begin{document}
\vspace*{0.35in}

\begin{flushleft}
{\Large
\textbf\newline{Monte Carlo simulations of 2D flat-sheet membrane filters \\ for constant-pressure water purification}
}
\newline
\\
Abigail Rose Drumm\textsuperscript{*},
Francesca Bernardi\textsuperscript{*,$\dagger$},
\\
\bigskip
\bf{*} Worcester Polytechnic Institute, 100 Institute Road, Worcester, MA 01609
\\
\bigskip
$\dagger$ Corresponding author: fbernardi@wpi.edu

\end{flushleft}

\section*{Abstract}
Membrane filtration is widely used in water treatment to remove foulants from contaminated water. Foulant build-up on the membrane occludes the area open for fluid flow, which impairs the efficiency of the filtration operation by decreasing the flux through the membrane. Backwashing is a strategy to restore the membrane, wherein clean water is processed backward through the membrane to dislodge attached foulants. We develop a Monte Carlo model to simulate constant-pressure forward filtration and backwashing through dead-end, flat-sheet membranes, with membrane fouling dominated by intermediate blocking. We validate our model against real-world experiments {\color{black} conducted with different foulant types and concentrations and run under different filtration conditions.} {\color{black} Relying primarily on measurable physical parameters and employing easy-to-implement parameter fitting techniques as needed, we} show good agreement between experimental {\color{black} data} and numerical simulations. {\color{black} We extend these results to predict flux behavior in forward filtration and backwashing when foulant properties or filtration conditions are varied.} The newly developed model can be used to further investigate the impact of varying backwashing duration, frequency, and/or pressure on the rate of flux recovery.

\section{\label{sec:intro}Introduction}

Increases in population, industry, and pollution over the past century have exacerbated the long-extant problem of clean water scarcity \cite{article:WWDR:BorettiRosa2019, review:GlobalWaterScarcity:vanVliet2021, article:WaterScarcityRivers:Wang_etal_2024}. Water purification requires many stages of treatment, among which membrane filtration is favorable for its relatively low energy requirements and capital cost \cite{article:MembraneTechReview:EzugbeRathilal2020, review:MembraneFiltrationTechniques:Aziz_etal_2024}. This technology has been applied in many industries, including water treatment for wastewater \cite{review:WaterReuseHistory:Angelakis_etal_2018, article:CleanWastewater:Tortajada2020, article:WastewaterSustainability:Obaideen_etal_2022}, surface water \cite{review:SurfaceGroundwaterFiltration:VanderBruggen2003, review:SurfaceWaterFiltration:HoslettETAL2018, article:SurfaceWaterEgypt:MahmoudETAL2018}, and groundwater \cite{review:SurfaceGroundwaterFiltration:VanderBruggen2003, review:MembraneFiltrationTechniques:Aziz_etal_2024}; food and drink manufacturing \cite{article:MembraneFood:Sutherland2010,  article:Charcosset2021:MembraneFood};
and downstream processing in the pharmaceutical industry \cite{review:MembranePharma:McKinnonAvis1993,article:vanReis2001:MembraneBiotech, article:BiopharmFiltration:Challener2024, article:BiopharmFiltrationVirus:QiETAL2025}.

Membrane filtration {\color{black} operates by two main mechanisms: (1) physiochemical interactions between foulants and the membrane and (2) sieving \cite{article:CombinedFouling:SanaeiCummings2019, book:BenjaminLawler:WaterQualityEng}. Our focus is on membrane microfiltration of} fluid fouled by suspended solids, bacteria, and other undesired materials {\color{black} that are larger than the membrane pores} \cite{article:Anis2019:MicrofiltrationReview}. {\color{black} Thus, here sieving is the dominant mechanism, and we split} membrane filtration processes {\color{black}into} two repeated steps: forward filtration (FF) followed by backwashing (BW). In forward filtration with dead-end, flat-sheet membranes, the fluid is pushed perpendicularly through the membrane and foulants accumulate directly onto its surface. The result is a phenomenon known as membrane fouling, wherein the membrane pores become increasingly blocked \cite{article:filtration:Tanudjaja_etal:2022, article:MembraneBlockingFluxDecline:BowenCalvoHernandez1995}. When filtration is run under constant-pressure conditions, membrane fouling results in considerable flux decline, which, in the long run, impairs the efficiency of and decreases the yield from filtration procedures \cite{article:ConstantPressureFiltration:HermansBredee1936, article:ConstantPressureFiltration:Hermia1982}. To counteract this decline, operators employ ``backwashing,'' running clean water backward through the membrane to dislodge foulants that have attached to it. This process recovers some, though not all, of the flux lost during forward filtration.

Backwashing presents a potentially significant loss in filtration output due to clean water waste and filtration downtime. In the field, backwashing frequency and duration are generally chosen empirically, based on expensive on-site pilot-scale studies, membrane manufacturer recommendations, individual facility protocol, and water quality \cite{web:BackwashingGuideline:Allgeier2005, web:BackwashingTechBrief:Satterfield2005}. To address this productivity loss, significant efforts are being employed to optimize the frequency and duration of backwashing so as to minimize the amount of clean water sacrificed while also maintaining desired water production volume and membrane integrity. Existing research has used optimal control theory \cite{article:OptimalControlBackwash:CoganChellam2014, article:OptimalControlBackwash:Kalboussi_etal_2018}, scheduling algorithms and control systems \cite{article:OptimalBackwashControl:Smith_etal_2005, article:OptimalBackwashControl:Smith_etal_2006, article:OptimalBackwashSchedule:Jepsen_etal_2019}, machine learning models \cite{article:OptimalBackwashML:Zhang_etal_2020, review:MembraneBackwashML:DansawadETAL2023}, and physical experimentation \cite{article:OptimalBackwashExperiment:Enten_etal_2020, article:OptimalBackwashHollowExp:SangrolaETAL2020, article:OptimalBackwashChemicals:KaykhaiiETAL2023} to optimize backwashing protocols.

In the current study, we propose a Monte Carlo method-based model to simulate forward filtration, fouling, and backwashing in dead-end, constant-pressure membrane filtration to capture flux decline and recovery. Monte Carlo methods have been previously employed to model pore blocking \cite{article:MonteCarloPoreCake:ChenKim2008, article:MonteCarloPoreBlocking:ChenHuKim2011}, cake formation \cite{article:MonteCarloCake:KimHoek2002, article:MonteCarloPoreCake:ChenKim2008, article:MonteCarloDeposition:LiVuKim2009, article:MonteCarloCake:Guan_etal_2017, article:PhysRevMCpaper:Lebovka_etal_2022}, and general particle deposition and membrane fouling \cite{article:MonteCarloCake:ChenElimelechKim2005, article:MonteCarloDeposition:LiVuKim2009, article:MonteCarloFoulantAttachment:Liu_etal_2020, article:MonteCarloDeposition:Wu_etal_2022}, but not backwashing or repeated FF-BW cycles. Using ideas from stochastic processes and classical membrane filtration theory, we successfully simulate the rate of flux decline, as well as post-backwashing recovery, observed in real-world experiments involving a variety of foulant types, membrane pore sizes, and transmembrane pressures. To improve the usability of the model, we incorporate a simple method to fit unknown physical or model parameters. Finally, we use the model to predict flux decline in untested filtration scenarios.

\section{\label{sec:methods}Methods}

\subsection{Domain}\label{subsec:domain}

As done in earlier numerical experiments for membrane fouling \cite{article:2DStochasticMembrane:Wessling2001, article:StokesletsDeadEndMicrofiltration:CoganChellam2008, article:PhysRevMCpaper:Lebovka_etal_2022}, we model the transport of fluid and suspended foulants in a dead-end filter by simplifying our system to a two-dimensional parallel plates domain. We assume radial and axial symmetry in the physical 3D cylindrical filter, so a 2D domain is sufficient to capture real-world foulant dynamics.

\begin{figure}[H]
        \subfloat[]{%
        \includegraphics[width=0.75\linewidth]{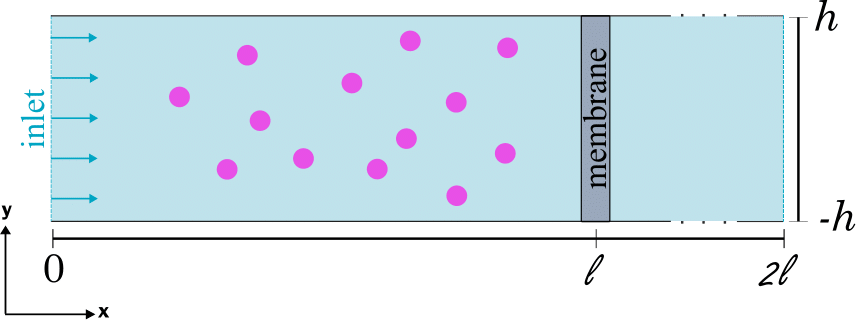}%
            \label{fig:domain}%
        } \hfill
        \subfloat[]{%
         \includegraphics[width=0.25\linewidth]{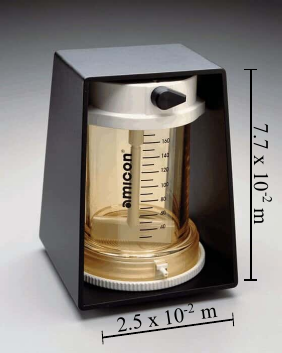}%
            \label{fig:millipore}%
        }
        \caption{(a) Schematic of modeled 2D domain. The channel presents solid horizontal boundaries at the top and bottom ($y=h$ and $y=-h$, respectively); it is open at the inlet ($x=0$) and outlet ($x=2\ell$), and includes a cross-sectional flat-sheet porous membrane at ($x = \ell$). The membrane is displayed here with some thickness for visualization purposes but is modeled as a one-dimensional barrier \cite{article:PartialDepositionMonteCarlo:VeerapaneniWiesner1994}. 
        (b) EMD Millipore 5121 Amicon Stirred Cell Model, produced by the Merck Group. This is the filter used in all experiments referenced in Section \ref{sec:results_disc} (without the stirrer). The photo of the stirred cell is from Millipore \cite{misc:Model8010:Millipore}, edited to include the scale bar.}
    \end{figure}

We consider a rectangular domain with dimensions $(x,y) = [0,2\ell] \times [-h,h]$ and a flat-sheet membrane at position $x = \ell$, as shown in Figure \ref{fig:domain}. The parallel plates domain represents solid top and bottom boundaries, mimicking the side walls of a filter, while the inlet at $x=0$ and the outlet at at $x=2\ell$ are open to allow foulants and water to enter the system.

We assume constant-pressure operations inducing a unidirectional flow in the positive $x$ direction. Following the set-up of \cite{article:StokesletsDeadEndMicrofiltration:CoganChellam2008, article:UnsteadyFiltrationPlug:DavydovaETAL2013}, we additionally assume that the distance $\ell$ between inlet and membrane is not sufficient for the flow within the channel to fully develop, so the system operates under a plug flow rather than parabolic flow condition; that is, the flow profile along each infinitesimal longitudinal slice of the rectangular domain has the same velocity $\forall y \in (-h, h)$. This assumption is consistent with typical filter dimensions and the experimental investigations to which we compare for benchmarking \cite{article:ConstantPressureDeadEnd:XuChellam2005, article:LakeHoustonCycles:GamageChellam2014}. For instance, the Amicon stirred cell shown  in Figure \ref{fig:millipore} (EMD Millipore 5121) has length $7.7 \times 10^{-2} \, \SI{}{\meter}$ and radius
 $1.25 \times 10^{-2} \, \SI{}{\meter}$ \cite{misc:Model8010:Millipore}; this is the filter utilized for all the experiments we refer to in Section \ref{sec:results_disc}. The membrane used throughout the experiments is $2.0 \times 10^{-5} \, \SI{}{\meter}$ thick with a radius matching that of the filter. We are able to model the membrane as a one-dimensional barrier since its width is much smaller than its radius and we assume that foulants can only settle on the membrane surface, i.e., cannot squeeze into the membrane pores; we discuss this further in the next section.

\subsection{Forward Filtration}

Our model for dead-end filtration captures forward filtration, membrane fouling, and backwashing, as mentioned in Section \ref{sec:intro}. We model forward filtration in two parts: the advection-diffusion of the foulants in the forward flow (Section \ref{subsubsec:FF-AD}) and the adhesion of the foulants to the membrane surface and consequent flux decline (Section \ref{subsubsec:FF-graft}).

\subsubsection{Advection-Diffusion}\label{subsubsec:FF-AD}

During forward filtration (FF), foulant particles undergo an advection-diffusion process given by:
\begin{equation}\label{eqn:adv_diff}
    \frac{\partial C}{\partial t} = - \vec{u}(t) \cdot \nabla C + \kappa \nabla^2 C,
\end{equation}
where $C(\vec{x},t)$ is the concentration of foulants in the feed water (cells \SI{}{\per \meter \cubed}), $\vec{x} = (x,y)$ is the position, $t$ is time (\SI{}{\second}), $\vec{u}(t) = (u_1(t), u_2(t))$ is the flux ((\SI{}{\meter \cubed \per \second})(\SI{}{ \per \meter \squared}) = \SI{}{\meter \per \second}), and $\kappa$ is the foulants' diffusion coefficient (\SI{}{\meter \squared \per \second}).

We take advantage of the well-known connection between diffusive and stochastic processes \cite{article:SDE_advdiff:AnceyETAL2015, book:SDE_advdiff:Thygesen2023} to model our system dynamics, as previously done in \cite{article:TracerSDE:ONaraigh2011, report:NumAnalysisSDE:ErhelETAL2014, article:SquaringCircle:Aminian_etal_2015, article:BoundariesMicrofluidics:Aminian_etal_2016, article:DiffusionTracers:Aminian_etal_2018}. Let $B(t)$ be a Brownian motion in time $t$ with drift parameter $\mu$ and variance parameter $2\kappa = \sigma^2$, a random function satisfying three properties:
\begin{enumerate}
    \item $B(t) - B(0) \sim \mathcal{N}(\mu t, \sigma^2 t)$,
    \item $B(t_2) - B(t_1) = B(t_2 - t_1) - B(0)$, $t_2 \geq t_1$,
    \item $B(t) = \mu t + \sigma W(t),$
\end{enumerate}
where $\mathcal{N}(\mu t, \sigma^2 t)$ is a normal distribution with mean $\mu t$ and variance $\sigma^2 t$, and $W(t)$ is a standard Brownian motion \cite{book:BrownianMotion:MortesPeres2008, book:StochasticCalculus:Lawler2014}. We define:
\begin{equation}\label{eqn:conc_pdf}
    C(\vec{x},t) = \lim_{\Delta \vec{x} \rightarrow 0} \frac{\mathbb{P}(\vec{x} \leq B(t) \leq \vec{x} + \Delta \vec{x})}{\Delta \vec{x}},
\end{equation}
to be a probability density function describing the probability that a particle is at a given position $\vec{x}$ at a given time $t$. The concentration definition in (\ref{eqn:conc_pdf}) satisfies
equation (\ref{eqn:adv_diff}) with $\kappa = \sigma^2/2.$

By generating many independent sample paths of a Brownian motion and approximating their probability density, we can approach the solution $C(\vec{x},t)$ for equation (\ref{eqn:adv_diff}) \cite{book:SDE_advdiff:Thygesen2023}. Hence, we track the spatial positions of foulant particles moving randomly in space within the filter domain. By incorporating our assumptions of unidirectional plug flow with some flux $u_1(t) = J(t)$ (\SI{}{\meter \per \second}) (and $u_2(t) = 0$), we obtain the following stochastic differential equation (SDE) for the foulant motion:
\begin{equation}\label{eqn:SDE_advdiff}
    d\vec{X}(t) = \vec{u} \, dt + \sqrt{2 \kappa} \, d \vec{W}(t).
\end{equation}
Here, $d\vec{X}(t) = (dX(t), dY(t))$ represents displacement in the axial and traverse directions, respectively; $\vec{u}(t) = (J(t), 0)$; and $dW(t) = (dW_1(t), dW_2(t)),$ where $dW_1(t)$ and $dW_2(t)$ are independent, standard Brownian motions (with unit $\SI{}{\second^{1/2}}$) that simulate random molecular diffusion in the horizontal and vertical directions, respectively.

Our Monte Carlo method builds on equation (\ref{eqn:SDE_advdiff}). We define the concentration of a given foulant in the contaminated water being filtered as $F_{\textnormal{conc}}$ (cells \SI{}{\per \meter \cubed}) and let $F_0$ be the number of foulants in the channel at time $t = 0$ based on $F_{\textnormal{conc}}$ \cite{article:PhysRevMCpaper:Lebovka_etal_2022}. We compute $n_{\textnormal{exp}}$ (cell~\SI{}{\per \second}) as the number of foulants that enter the domain at every time step based on the initial flux, $J_0$ (\SI{}{\meter \per \second}). 

Each individual foulant particle in the domain moves according to the dynamics of equation (\ref{eqn:SDE_advdiff}). At each time step, a foulant moves forward in the positive $x$ direction by a distance $J(t) dt$ (in \SI{}{\meter}) under the influence of the background advective plug flow. In addition, it randomly moves ``north,'' ``east,'' ``south,'' or ``west'' $\sqrt{2 \kappa t}$ units (in \SI{}{\meter}) due to diffusion. To ensure that foulants do not leave the channel, we impose billiard-like reflections at the top and bottom solid boundaries ($y=h$ and $y=-h$, respectively) \cite{article:SilaneDiffMicroMachines:Howard_etal_2021}. 

\subsubsection{Foulant grafting \& flux decline}\label{subsubsec:FF-graft}

We assume complete contaminant rejection during filtration due to size, i.e., we only model the behavior of foulants larger than the membrane pores and unable to pass through them. This implies that the membrane does not experience ``standard blocking,'' with foulants aggregating along the interior walls of the pores; rather, foulant accumulate on its surface \cite{article:ReviewPoreBlocking:Iritani2013}. This, in combination with the fact that the membrane thickness is much smaller than its radius, allows us to model the membrane as a one-dimensional barrier.

When a foulant particle approaches the membrane, it either grafts to its surface or it bounces back into the channel flow. Whether each foulant particle attaches or bounces back is determined probabilistically. Any given foulant has an associated probability of grafting to the membrane, $P_{gr}$, that is uniformly distributed between 0 and some maximum probably of grafting $p_{gr}$, i.e., $P_{gr} \in \textnormal{Unif}(0, p_{gr})$. The probability $P_{gr}$ may be additionally adjusted by the particle size and whether other foulants are already stuck on the area of the membrane where a foulant would attach. We allow multiple foulants to occupy position\textcolor{black}{s} on the membrane \textcolor{black}{in very close proximity} to conceptualize foulant \textcolor{black}{overlap.}

We model fouling dynamics as dominated by ``intermediate pore blocking'' where foulants \textcolor{black}{can overlap} and cover both the membrane surface and the membrane pores due to their size \cite{article:MembraneBlockingFluxDecline:BowenCalvoHernandez1995, article:ConstantPressureDeadEnd:XuChellam2005}; this creates a decline in flux over the course of forward filtration. Experimental analysis suggests that membrane filtration transitions from the intermediate pore blocking regime to the ``cake filtration'' regime when 10 -- 20\% of the membrane area remains open  \cite{article:ConstantPressureDeadEnd:XuChellam2005}. The rate of membrane area coverage depends on the contamination of the water, the membrane pore sizes, and the duration of filtration. At the concentration levels, pore sizes, and filtration runtimes modeled in the experimental set-ups we compare to in Section \ref{subsec:FF_validation}, this degree of membrane area coverage is not achieved and, thus, the cake formation phase of filtration is not reached \cite{article:ConstantPressureDeadEnd:XuChellam2005}.

The established equation for flux decline due to intermediate blocking is \cite{article:ConstantPressureFiltration:Hermia1982, article:ReviewPoreBlocking:Iritani2013}:
\begin{equation}\label{eqn:intermediate_blocking_flux}
    J(t) = \frac{J_0}{1 + J_0 K_i t}.
\end{equation}
Here, $J_0$ is the initial flux (\SI{}{\meter \per \second}), $K_i$ is the membrane areal coverage per unit volume of feed filtered (\SI{}{\meter \squared}$\cdot$\SI{}{\per \meter \cubed} = \SI{}{\per \meter}), and $t$ is time (\SI{}{\second}).

We adapt equation (\ref{eqn:intermediate_blocking_flux}) to model the flux decline in our framework by relating numerical quantities to the physical parameters describing time-dependent areal coverage in real-world filtration processes. First, we denote $F_{\textnormal{gr}}(t)$ (cells) to be the number of foulants that have grafted on the membrane at time $t$, and $F_{\textnormal{area}}$ (\SI{}{\meter \squared} cell$^{-1}$ ) to be the lengthwise cross-sectional area of a single foulant particle. Then $F_{\textnormal{gr}}(t) \cdot F_{\textnormal{area}}$ gives the areal coverage of the simulated membrane at time $t$ due to the accumulation of foulants. \textcolor{black}{We note that such areal coverage is likely an overestimation due to partial overlap of some grafted foulants during intermediate blocking, which is permitted in the model, as described earlier.}

As noted previously, we have $n_{\textnormal{exp}}$ particles entering the channel at every time step, where $n_{\textnormal{exp}}$ is a function of the foulant concentration and the initial flux. In practice, a typical $n_{\textnormal{exp}}$ would be of the order of $10^4$. For computational efficiency in the simulation (where we would have to introduce $10^4$ at each time step), we scale $n_{\textnormal{exp}}$ by $10^{\textnormal{s}}$ for some $s \in \mathbb{Z}^+$, and define the actual number of particles we introduce at each time step as $n = n_{\textnormal{exp}}/10^{s}$ (cell \SI{}{\per \second}). To compute the actual number of foulants that would attach to the membrane in the physical, real-world experiment, we scale $F_{\textnormal{gr}}(t)$ by $F_{\textnormal{conc}}/n$.
Altogether, we model the areal membrane coverage due to the attached foulants at time $t$ per unit volume filtered (\SI{}{\per \meter \second}) as:
\begin{equation}\label{eqn:Monte_Carlo_sigma}
    K_i t = \frac{F_{\textnormal{conc}}}{n} F_{\textnormal{gr}}(t) F_{\textnormal{area}}A_{\textnormal{adj}}.
\end{equation}
We add a non-dimensional model parameter, $A_{\textnormal{adj}}$, to incorporate the influence of any unaccounted for parameters, such as membrane pore size, on flux decline.
Finally, we define the flux expression in our Monte Carlo simulations to be:
\begin{equation}\label{eqn:sim_intermediate_flux}
    \displaystyle{J(t) = \frac{J_0}{1 + J_0  (F_{\textnormal{conc}}/n)  F_{\textnormal{gr}}(t)   F_{\textnormal{area}}A_{\textnormal{adj}}}.}
\end{equation}
This formula describes how the flux $J(t)$ through the channel is affected by the foulant accumulation on the membrane, quantified by $F_{\textnormal{gr}}(t)$. Over time, as foulants amass, we expect the flux to be monotonically decreasing and concave up \cite{article:MembraneBlockingFluxDecline:BowenCalvoHernandez1995,article:ConstantPressureDeadEnd:XuChellam2005,article:StokesletsDeadEndMicrofiltration:CoganChellam2008, article:BacterialFoulingNumerics:CoganChellam2009}.

\subsubsection{Nondimensionalization}\label{subsec:nondim}

We developed our model in dimensional coordinates to enable straightforward connection with and application to physical, real-world experiments. Now we introduce a nondimensional form of the stochastic advection-diffusion equation (\ref{eqn:SDE_advdiff}) driving the particle movement during forward filtration. We denote dimensionless variables by carets as:
$$\hat{X} = \frac{X}{h}, \quad  \hat{Y} = \frac{Y}{h}, \quad \hat{J} = \frac{J}{J_0}, \quad \hat{t} = \frac{t}{h^2/\kappa}.$$
We nondimensionalize the spatial terms by the channel half-width $h$ (\SI{}{\meter}), the flux by the initial forward filtration flux $J_0$ (\SI{}{\meter \per \second}), and time by the diffusion timescale $h^2/\kappa$ (\SI{}{\second}), where $\kappa$ is the foulants' diffusion coefficient (\SI{}{\meter \squared \per \second}).

The nondimensional stochastic advection-diffusion equation is:
\begin{equation}
\label{eqn:SDE_advdiff_nondim}
    d\hat{\vec{X}}(\hat{t}) = \Pen \cdot \hat{\vec{u}}(t) dt + \sqrt{2} d\hat{\vec{W}}(t), 
\end{equation}
where $\Pen = (J_0 \cdot h)/\kappa$ is the P{\'e}clet number, a nondimensional parameter indicating the relative importance of advection and diffusion, and $\hat{\vec{u}}(t)$ and $d\hat{\vec{W}}(t)$ are defined as in equation (\ref{eqn:SDE_advdiff}). Since only diffusion (no advection) occurs in the $y$-direction, we have $\hat{\vec{u}} = (\hat{J}(t), 0)$ and $\Pen$ does not appear in the transverse equation. When Pe is small, the diffusive term $d\hat{W_1}(\hat{t})$ plays a more substantial role in the translation of particles in the flow. By contrast, when Pe is large, advection dominates and the flux curves will appear smoother.

In microfiltration, the advective transport often dominates over the diffusive transport, resulting in relatively large $\Pen$ values, typically ranging between $10$ and $10^5$ \cite{article:PecletRange:StoneETAL2004, chapter:MicrofluidicTheory:Microfluidics:ManzETAL2020}.
In the experimental studies we compare to, $J_0 \sim 10^{-5} - 10^{-4} \, \SI{}{\meter \per \second}$, $h \sim 10^{-2} \, \SI{}{\meter}$, and $\kappa \sim 10^{-10} \, \SI{}{\meter \squared \per \second}$, so $\Pen \sim 10^3 - 10^4$ \cite{article:ConstantPressureDeadEnd:XuChellam2005, article:LakeHoustonCycles:GamageChellam2014}. Throughout this work, we use $\kappa = 10^{-10} \, \SI{}{\meter \squared \per \second}$, which is the typical magnitude for the diffusion coefficient of bacteria such as those in the benchmarking studies of Section \ref{subsec:FF_validation} \cite{article:BacteriaDiffCoeff:Kim1996,article:BacteriaDiffCoeff:Licata2016}.

\subsection{Backwashing}\label{subsec:BW}

After the forward filtration phase, we introduce backwashing (BW) to clean the membrane and thereby restore some of the flux lost by membrane fouling. Not all of the flux lost during forward filtration is recoverable during backwashing due to the presence of untouched or irreversibly attached foulants \cite{article:IrreversibleFouling:KimuraETAL2006, article:RemainingFouling:RemizeETAL2010,article:LakeHoustonCycles:GamageChellam2014,article:IrreversibleFouling:GuptaChellam2022}. While we currently do not differentiate between types of foulants, our model does capture these leftover particles.

In a modeling sense, the backwashing mechanism consists of sending fluid particles (with no foulants) towards the membrane in a direction opposite that of forward filtration, from outlet towards inlet in Figure \ref{fig:domain}. When a fluid particle arrives at the membrane at a location where a foulant is attached on the other side, it can either dislodge it, flow through the membrane without any impact on the membrane fouling (if the pore is not completely blocked), or bounce back (if the pore is completely blocked). As done with grafting in forward filtration, we assign to each attached foulant a probability of removal, $P_{rem}$, that is uniformly distributed between 0 and some maximum probably of removal $p_{rem}$, i.e., $P_{rem} \in \textnormal{Unif}(0, p_{rem})$. 

We expect the maximum probability of removal to be much smaller than the maximum probability of grafting ($p_{rem} \ll p_{gr}$) based on the premise that due to the number of membrane-foulant and foulant-foulant interactions involved that resist detachment during the BW phase, it takes more encounters of consecutive water parcels to dislodge an attached foulant than it takes an individual foulant to graft to the membrane  \cite{article:DLVOAttachment:Oliveria1997, article:EDLVO_Fouling:FuETAL2018, review:MembraneFoulantInteractionsReview:XuETAL2020}.

Due to the high computational cost that the tracking of individual fluid particles would present, we simplify the simulation of the backwashing mechanism. We do not track individual fluid particles as they approach the membrane but rather assume that a fixed number reaches the membrane at each time step. Then, we randomly select a corresponding number of attached foulants as candidates for removal at each time step and probabilistically determine whether these candidates are successfully removed. 

The number of foulants selected for potential removal is determined similarly to the input $n$ used in forward filtration. We define $n_{\textnormal{BW}}$ as the number of particle candidates for removal at each time step, and set it equal to the number of fluid parcels that reach the membrane every second. The volumes of these modeled parcels match those of the foulants. 
We choose to consider the number of fluid parcels (rather than molecules) reaching the membrane based on the assumption that, to remove a foulant of a given size, we would need a volume of fluid that is of comparable size. Moreover, the number of water molecules in the volumes filtered would far exceed the computational bounds of our software. In this manuscript, we will work with foulants with cross-sectional areas varying between $10^{-13}$ and $10^{-9} \, \SI{}{\meter \squared}$.

Flux recovery in backwashing is accomplished by the successful removal of foulants attached to the membrane. Equation (\ref{eqn:sim_intermediate_flux}) applies here as well; the impact of backwashing is reflected in the flux decline formula through a gradual decrease in the number of foulants grafted at the membrane, $F_{\textnormal{gr}}(t)$. As noted, in physical filtration operations flux recovery is not total, as some foulants remain attached to the membrane \cite{article:RemainingFouling:RemizeETAL2010,article:IrreversibleFouling:GuptaChellam2022}; this effect is captured by our backwashing simulations.

\section{\label{sec:results_disc}Results \& Discussion}

We show here the validity of our Monte Carlo-based numerical model for capturing (A) the trends of membrane fouling and flux decline observed in real-world experiments; and (B) the partial flux recovery achieved over repeated cycles of forward filtration (FF) and backwashing (BW). The model captures membrane fouling and flux behavior in filtration scenarios with varying transmembrane pressure (TMP), foulant sizes, and pore sizes. 

In Section \ref{subsubsec:benchmarking}, we compare simulation results for forward filtration to experiments that use lab-made feed water with a single type of bacteria \cite{article:ConstantPressureDeadEnd:XuChellam2005}. In Section \ref{subsec:FF-BW_validation}, we compare simulations for FF-BW cycling to experiments that use lake water samples \cite{article:LakeHoustonCycles:GamageChellam2014}. We also use our benchmarked numerical model to predict the flux decline in filtration scenarios not explored in existing work (Section \ref{subsubsec:extensions}).

For all of the simulations, we fix the maximum probability of grafting to be $p_{gr} = 1$ so that, on average, it takes two collisions with the membrane for a foulant to graft. We fit the nondimensional model parameter $A_{\textnormal{adj}}$ to the experimental data by minimizing the normalized residual between the normalized flux of the simulations, $\tilde{J}(t) = J(t)/J_0,$ and the normalized flux of the experimental data, $\tilde{J}_{\textnormal{exp}}(t) = J_{\textnormal{exp}}(t)/J_0,$ across all recorded times. We define the normalized residual as:
\begin{equation}\label{eqn:res}
    \textnormal{Normalized Residual} = \sum_{t} \left(\tilde{J}(t) - \tilde{J}_{\textnormal{exp}}(t) \right)^2,
\end{equation}
and implement a parameter estimation routine using MATLAB 2024a's built-in \texttt{fminsearch} derivative-free optimization method, which uses the Nelder-Mead simplex method \cite{article:NelderMeadOriginal:1965, article:NelderMead:LagariasETAL1998}. We apply this same method to fit more than one parameter simultaneously in Section \ref{subsec:FF-BW_validation}.

\subsection{Forward filtration operations}\label{subsec:FF_validation}

\subsubsection{Model benchmarking}\label{subsubsec:benchmarking}

\begin{figure}[H]
\centering
\includegraphics[width=0.75\linewidth]{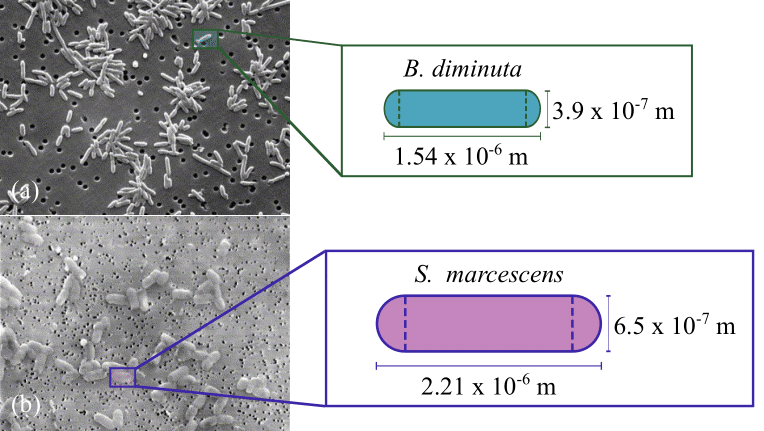}

\caption{Schematic of modeled bacteria species (a) \textit{B.~diminuta} and (b) \textit{S.~marcescens}. Water samples presenting fixed concentrations of one of these species of bacteria at a time are used to benchmark forward filtration operations, as shown in Figures \ref{subfig:bdimlow}-\ref{subfig:smarchigh}. The average length (from end to end) and diameter dimensions of each bacteria species are reported in \cite{article:ConstantPressureDeadEnd:XuChellam2005}. Scanning electron microscope (SEM) images of the bacteria are from \cite{article:ConstantFluxDeadEnd:ChellamXu2006}, used with permission from Elsevier.}\label{fig:bacteria}

\end{figure}

We benchmark our forward filtration flux decline model by comparing simulation results to constant-pressure, dead-end filtration experiments of bacteria-fouled water conducted by Xu and Chellam with flat-sheet membrane filters \cite{article:ConstantPressureDeadEnd:XuChellam2005, article:StokesletsDeadEndMicrofiltration:CoganChellam2008}. In these bench-top investigations, feed water containing a prescribed concentration of a single rod-shaped bacteria species, either \textit{Brevundimonas diminuta} or \textit{Serratia marcescens} (from now on, \textit{B.~diminuta} and \textit{S.~marcescens}, respectively), underwent constant-pressure, dead-end filtration through track-etched polycarbonate membrane filters \cite{article:ConstantPressureDeadEnd:XuChellam2005}. When undergoing advection-diffusion, the bacteria are treated as particles with no volume or shape. When grafted onto the membrane, we model the bacteria as pill-shaped organisms with cylindrical middles and hemispherical caps, as pictured in Figure \ref{fig:bacteria}.

The filtration experiments were completed in commercial filters set up without stirrers (see Figure \ref{fig:millipore}) designed to accommodate membranes with an effective area of $4.1 \times 10^{-4}$ \SI{}{\meter \squared} \cite{article:ConstantPressureDeadEnd:XuChellam2005}. As mentioned, the cylindrical filters have length $7.7 \times 10^{-2} \, \SI{}{\meter}$ and radius $1.25 \times 10^{-2} \, \SI{}{\meter}$, which we set to be $\ell$ and $h$, respectively, for the 2D model domain. 

\begin{figure}[H]
    \centering
    \includegraphics[width=0.65\linewidth]{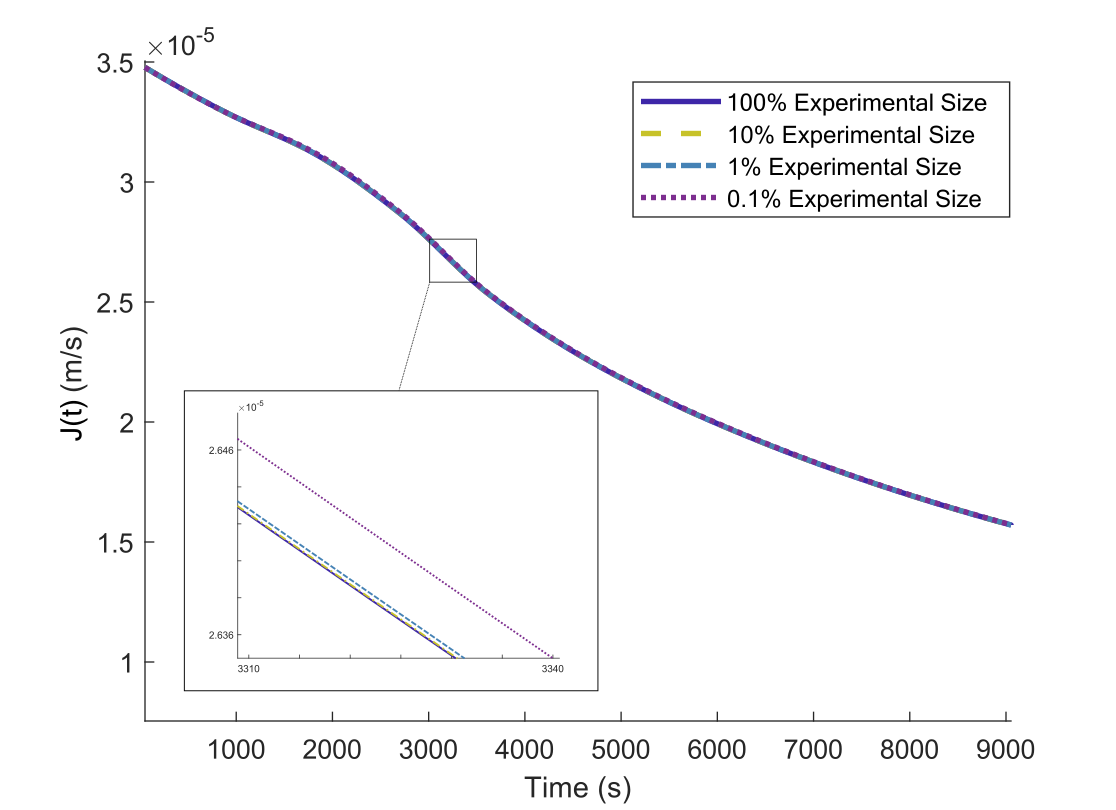}%
        \caption{Comparison of simulations run at 0.1\%, 1\%, 10\%, and 100\% of the original experimental bacterial concentration size of order $10^{12} \, \textnormal{cells } \SI{}{\per \meter \cubed}$. We include an inset of a zoomed in portion of the full figure to better illustrate the minuscule differences between the curves. We use experimental parameters from Figure \ref{subfig:smarclow} for the simulations. The runtime and the normalized residuals relative to the experiments run at full experimental size are given in Table \ref{tab:scale-stats}.}
\label{fig:scale}
\end{figure}

\begin{figure*}
\captionsetup[subfigure]{justification=centering} 
        \subfloat[]{%
            \includegraphics[width=0.48\linewidth]{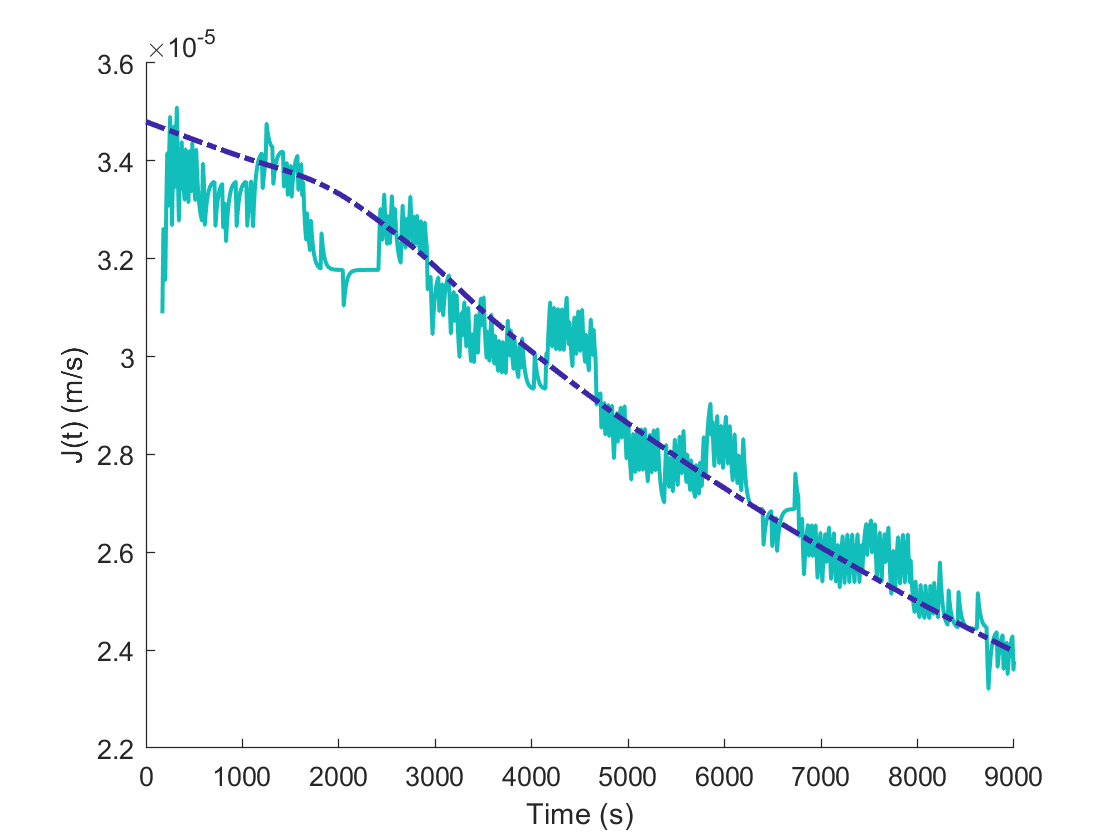}%
            \label{subfig:bdimlow}%
        }\hfill
        \subfloat[]{%
            \includegraphics[width=0.48\linewidth]{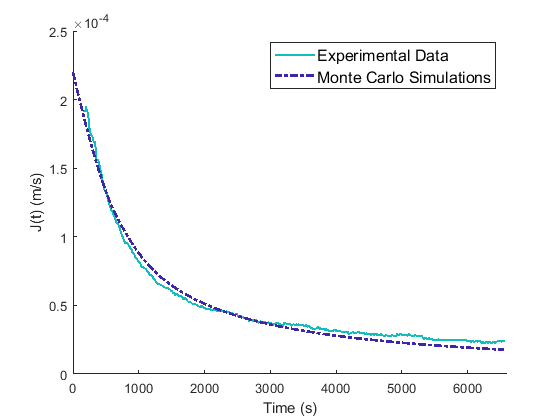}%
            \label{subfig:bdimhigh}%
        }\\
        \subfloat[]{%
            \includegraphics[width=0.48\linewidth]{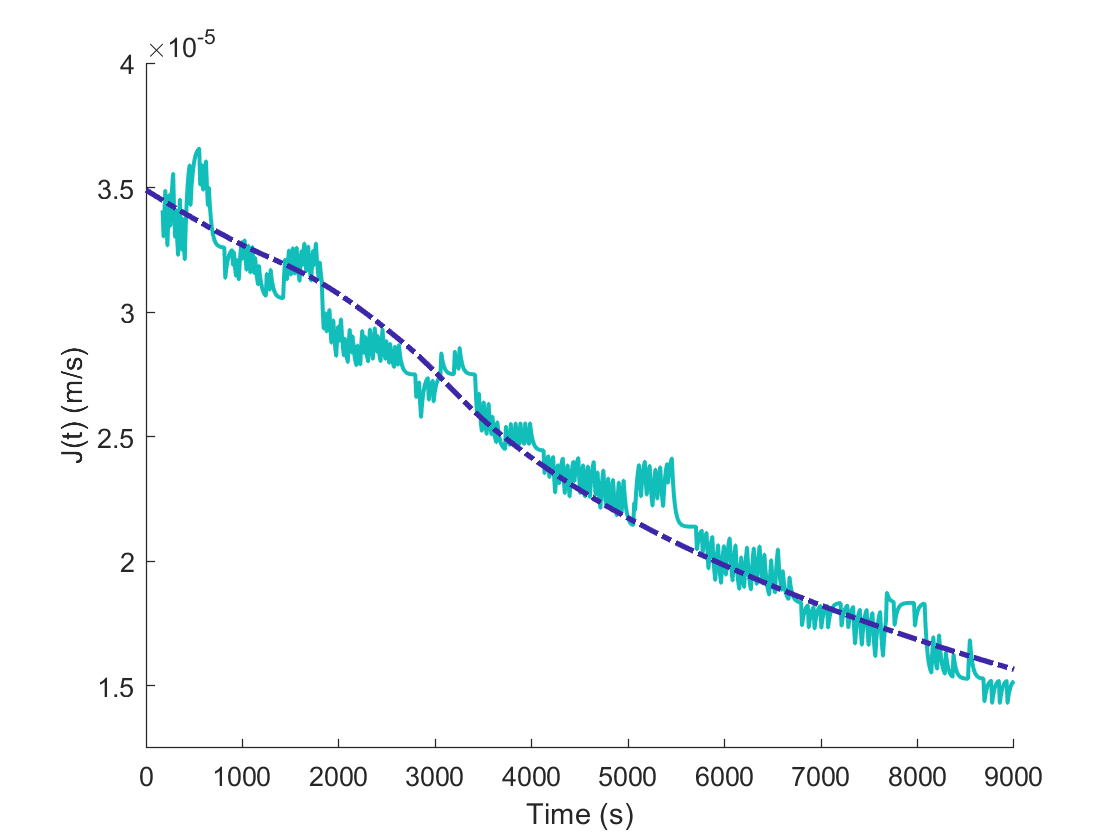}%
            \label{subfig:smarclow}%
        }\hfill
        \subfloat[]{%
            \includegraphics[width=0.48\linewidth]{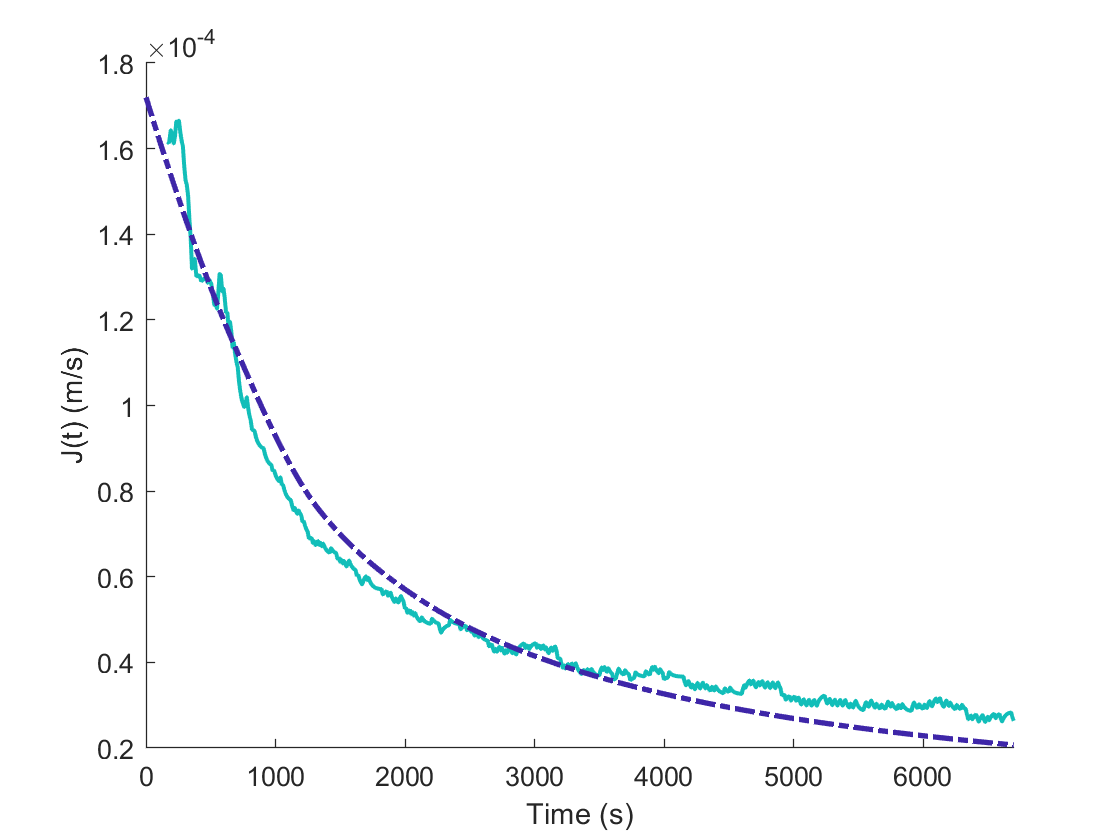}%
            \label{subfig:smarchigh}%
        }
        \caption{Forward filtration simulations. Each simulation curve is the average of 100 runs. Parameters for each figure are given in Table \ref{tab:simparams}.
        (a) Simulation of feed water with $2.86 \times 10^{12}$ cells~\SI{}{\per \meter \cubed} of \textit{B.~diminuta} filtered through membrane with 400-nm pores at a constant TMP of $35.852 \times 10^3 \, \SI{}{\pascal}$ (5.2 psi) and initial flux of $3.48 \times 10^{-5} \, \SI{}{\meter \per \second}$. Experimental data from Figure 4 of \cite{article:StokesletsDeadEndMicrofiltration:CoganChellam2008}.
        (b) Simulation of feed water with $1.53 \times 10^{13}$ cells~\SI{}{\per \meter \cubed} of \textit{B.~diminuta} filtered through membrane with 200-nm pores at a constant TMP of $33.577 \times 10^3 \, \SI{}{\pascal}$ (4.87 psi) and initial flux of $2.2 \times 10^{-4} \, \SI{}{\meter \per \second}$. Experimental data from Figure 3 of \cite{article:ConstantPressureDeadEnd:XuChellam2005}.
        (c) Simulation of feed water with $2.75 \times 10^{12}$~cells~\SI{}{\per \meter \cubed} \textit{S.~marcescens} filtered through membrane with 200-nm pores at a constant TMP of $27.165 \times 10^3 \, \SI{}{\pascal}$ (3.94 psi) and initial flux of $3.49 \times 10^{-5} \, \SI{}{\meter \per \second}$. Experimental data from Figure 7 of \cite{article:StokesletsDeadEndMicrofiltration:CoganChellam2008}.
        (d) Simulation of feed water with $5.49 \times 10^{12}$ cells~\SI{}{\per \meter \cubed} of \textit{S.~marcescens} filtered through membrane with 400-nm pores at a constant TMP of $28.958 \times 10^3 \, \SI{}{\pascal}$ (4.2 psi) and initial flux of $1.72 \times 10^{-4} \, \SI{}{\meter \per \second}$. Experimental data from Figure 3 of \cite{article:ConstantFluxDeadEnd:ChellamXu2006}.}
        \label{fig:ffsims}
    \end{figure*}

Each simulation in section \ref{sec:results_disc} is run at 0.1\% of the original size ($s=3$), rescaling the experimental foulant concentration by $10^3$. Test simulations are shown in Figure \ref{fig:scale} with runtime and residual calculations summarized in Table \ref{tab:scale-stats}; these confirm that, for experimental concentrations of the order of $10^{12}$, setting the number of particles to be 0.1\% ($s=3$) of the experimental size is not significantly different from setting it to be 1\%, 10\% ($s=2$ and $s=1$, respectively), or 100\% ($s=0$); by eye, the differences are imperceptible. We provide a zoomed in view as an inset in Figure \ref{fig:scale} for reference. For example, in Figure \ref{subfig:bdimlow}, the original experiment had a bacterial concentration of $2.86 \times 10^{12}$~cells~\SI{}{\per \meter \cubed}; the artificial concentration in the numerical experiments corresponding to 0.1\% of the experimental size is $F_{\textnormal{conc}} = 2.86 \times 10^{9}$~cells \SI{}{\per \meter \cubed}.

\begin{table}[H]
\centering
\caption{\label{tab:scale-stats}%
Summary of runtime for each simulation in Figure \ref{fig:scale} and normalized residual difference relative to full (100\%) experimental size.}
\begin{tabular}{lll}
\hline
 Simulation Size & Runtime (hrs) & Normalized Residual \\ \hline
   100\% & 48.81 & n/a \\
   10\% & 4.95 & $1.221 \times 10^{-7}$ \\
   1\% & 0.52 & $5.295 \times 10^{-6}$ \\
   0.1\% & 0.05 & $5.664 \times 10^{-4}$ \\
   \hline
\end{tabular}
\end{table}

The physical experiments in \cite{article:ConstantPressureDeadEnd:XuChellam2005} generally involved filtration of $5.0 \times 10^{-4} \, \SI{}{\meter \cubed}$ of water with foulant concentrations of the order of $10^{12} \, \textnormal{cells } \SI{}{\per \liter}$, totaling foulant numbers of $\mathcal{O}(10^8)$. We use 0.1\% of the experimental size in our simulations, so $\mathcal{O}(10^5)$ foulants total. 
We run 100 simulations with the same set of parameters and average the results from these independent realizations, to obtain $10^7$ sample paths, similar to \cite{article:SquaringCircle:Aminian_etal_2015, article:BoundariesMicrofluidics:Aminian_etal_2016, article:PassiveScalarTransportPipes:AminianETAL2018, article:SilaneDiffMicroMachines:Howard_etal_2021}.

Figure \ref{fig:ffsims} shows simulated flux decline curves compared against the experimental data from Xu and Chellam \cite{article:ConstantPressureDeadEnd:XuChellam2005}. Simulation parameters are provided in Table \ref{tab:simparams}. Each row in Figure \ref{fig:ffsims} is plotting results for the same bacteria species (top: \textit{B.~diminuta}; bottom: \textit{S.~marcescens}), while each column shows flux decline trends when the initial flux $J_0$ (\SI{}{\meter \per \second}) has the same order of magnitude. We include the corresponding P{\'e}clet number for each simulation; the Reynolds number is $\mathcal{O}(10^{-3} - 10^{-4})$.

Consistent with the predictions of membrane fouling dominated by intermediate blocking \cite{article:MembraneBlockingFluxDecline:BowenCalvoHernandez1995, article:ConstantPressureDeadEnd:XuChellam2005}, the flux decline curves are concave up {\color{black} on long time-scales. The curves in Figures \ref{subfig:bdimlow} and \ref{subfig:smarclow} are initially concave down due to the lower initial flux, $J_0$, employed in these experiments. Lower fluxes translate to slower arrival of foulants to the membrane and, thus, slower onset of the upward concavity characteristic of intermediate blocking.}

\begin{table}[H]
\caption{\label{tab:simparams}%
In-model parameters and variables used for simulations plotted in Figures \ref{subfig:bdimlow}-\ref{subfig:smarchigh}. The physical parameter values were derived or inferred from the original experiments \cite{article:ConstantPressureDeadEnd:XuChellam2005} or numerical experiments by the same research group \cite{article:StokesletsDeadEndMicrofiltration:CoganChellam2008, article:BacterialFoulingNumerics:CoganChellam2009}.}
\begin{tabular}{lllllll}
    \hline
  & Description (units) & Fig.~\ref{subfig:bdimlow} & Fig.~\ref{subfig:bdimhigh} & Fig.~\ref{subfig:smarclow} & Fig.~\ref{subfig:smarchigh} & Sources \\ \hline
   $J_0$ & Initial flux (\SI{}{\meter \per \second}) & $3.48 \times 10^{-5}$ & $2.2 \times 10^{-4}$ & $3.49 \times 10^{-5}$ & $1.72 \times 10^{-4}$ & \cite{article:ConstantPressureDeadEnd:XuChellam2005}, \cite{article:StokesletsDeadEndMicrofiltration:CoganChellam2008} \\

   $\Pen$ & P{\'e}clet number & 2175 & 13750 & 2181.25 & 10750 & - \\
   $F_{\textnormal{conc}}$ & Foulant concentration (cells \SI{}{\per \meter \cubed}) & $2.86 \times 10^{12}$ & $1.53 \times 10^{13}$ & $2.75 \times 10^{12}$ & $5.49 \times 10^{12}$ & \cite{article:ConstantPressureDeadEnd:XuChellam2005}, \cite{article:StokesletsDeadEndMicrofiltration:CoganChellam2008} \\
   $F_{\textnormal{area}}$ & Cross-sectional foulant area (\SI{}{\meter \squared} cell$^{-1}$) & $5.68 \times 10^{-13}$ & $5.68 \times 10^{-13}$ & $1.35 \times 10^{-12}$ & $1.35 \times 10^{-12}$ & \cite{article:ConstantPressureDeadEnd:XuChellam2005} \\
   $A_{\textnormal{adj}}$ & Model parameter (-) & 1.19 & 1.73 & 1.71 & 1.46 & Model fit \\
   $t_F$ & Forward filtration duration (\SI{}{\second}) & 9000 & 6560 & 9000 & 6700 & \cite{article:StokesletsDeadEndMicrofiltration:CoganChellam2008} \\
   $d_{\textnormal{pore}}$ & Pore diameter (\SI{}{\meter}) & $4.0 \times 10^{-7}$  & $2.0 \times 10^{-7}$ & $2.0 \times 10^{-7}$ & $4.0 \times 10^{-7}$ & \cite{article:ConstantPressureDeadEnd:XuChellam2005}, \cite{article:BacterialFoulingNumerics:CoganChellam2009} \\
   n & Input rate (cells~\SI{}{\per \second}) & 41 & 1381 & 40 & 388 & - \\
   \hline
\end{tabular}
\end{table}

As expected, experiments run with larger bacteria (Figure \ref{fig:ffsims}, bottom row) foul the membrane more quickly than those run with smaller bacteria (Figure \ref{fig:ffsims}, top row). Moreover, for the same bacteria type, a higher initial flux (Figure \ref{fig:ffsims}, right column) results in 
a steeper drop in flux over time. When comparing the results plotted in Figures \ref{subfig:bdimlow} and \ref{subfig:smarclow}, we observe greater overall flux decline in the latter, where the bacteria size is larger and the membrane pore size is smaller, though the experiments start with similar initial fluxes.
In Figures \ref{subfig:bdimhigh} and \ref{subfig:smarchigh}, we see that for sufficiently high initial fluxes (an order of magnitude larger than Figures \ref{subfig:bdimlow} and \ref{subfig:smarclow}) and large bacterial concentrations, the flux declines more much sharply.
Across the experiments, a higher initial flux corresponds to higher Pe number values, which indicate that advection dominates over diffusion; this is reflected in the smoother flux decline curves of Figures \ref{subfig:bdimhigh} and \ref{subfig:smarchigh} compared to Figures \ref{subfig:bdimlow} and \ref{subfig:smarclow}.

We rely primarily on five physical, real-world parameters to generate the flux decline simulations in Figure \ref{fig:ffsims}, as detailed in Table \ref{tab:simparams}; the only model parameter fit to the data is the adjustment $A_{\textnormal{adj}}.$ Using these five physical parameters and fitting one parameter, our model performs very well by matching experimental flux decline values with residual errors of 0.432 (Figure \ref{subfig:bdimlow}), 0.256 (Figure \ref{subfig:bdimhigh}), 0.835 (Figure \ref{subfig:smarclow}), and 0.676 (Figure \ref{subfig:smarchigh}).

\subsubsection{Model extensions}\label{subsubsec:extensions}

To further exhibit the versatility of the model, we vary the physical parameters of these experiments to approximate how changes in water composition or filtration conditions impact the measured flux decline. 

\begin{figure*}[h]
\captionsetup[subfigure]{justification=centering} 
        \subfloat[]{%
            \includegraphics[width=0.48\linewidth]{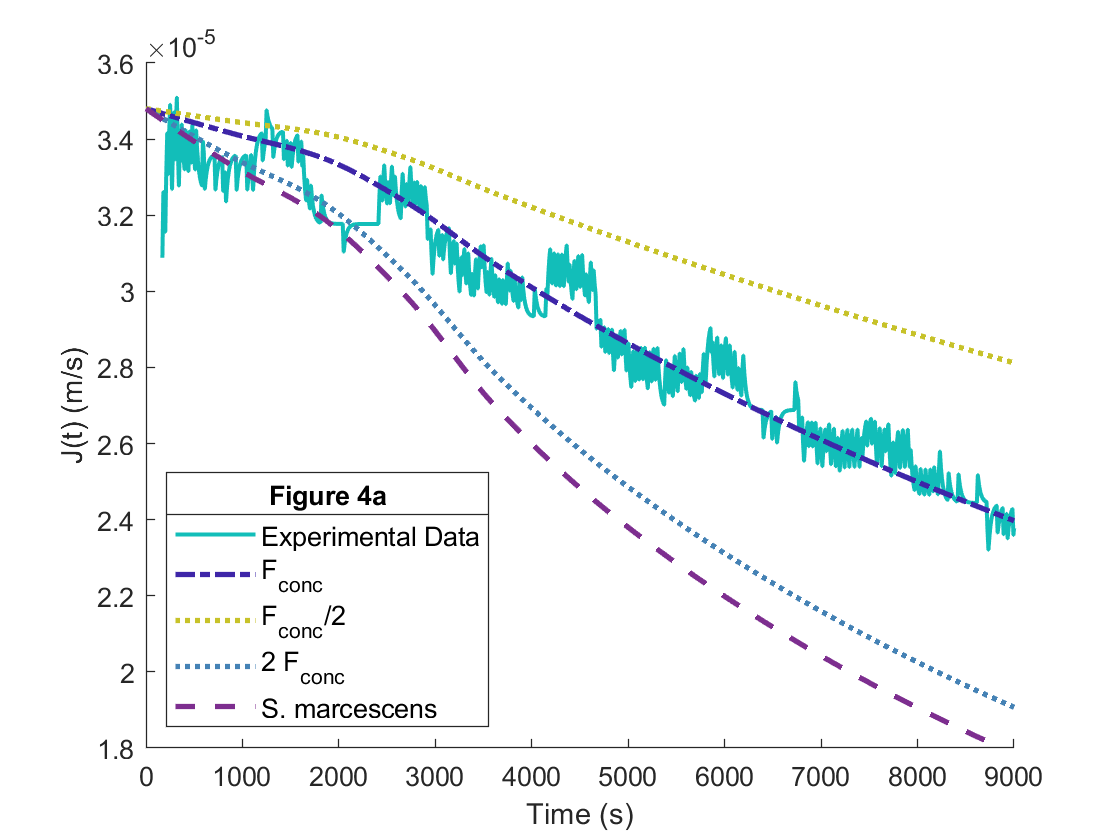}%
            \label{fig:bdimlow-conc}%
        }\hfill
        \subfloat[]{%
            \includegraphics[width=0.48\linewidth]{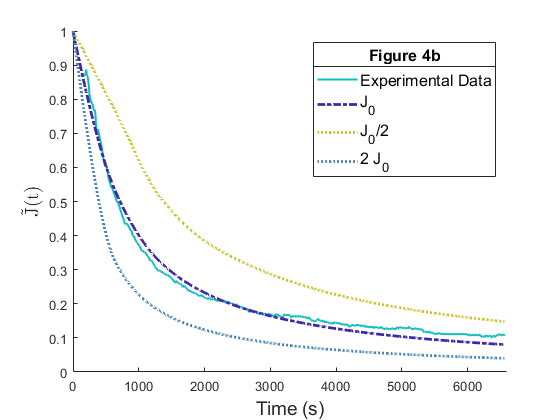}%
            \label{fig:bdimhigh-flux}%
        }
        \caption{Impact of varying a single parameter on the observed flux decline in one phase of forward filtration.
        (a) Comparing flux decline across simulations of feed water with \textit{B.~diminuta} filtered through a membrane with 400-nm pores at a base concentration of $2.86 \times 10^{12} \, \textnormal{cells} \, \SI{}{\per \meter \cubed}$ (\textcolor{Blue}{$- \cdot - \cdot$}), at half the base concentration (\textcolor{Dandelion}{$\cdot \cdot \cdot$}), and at double the base concentration using \textit{B.~diminuta} (\textcolor{NavyBlue}{$\cdot \cdot \cdot$}) and at base concentration using \textit{S.~marcescens} (\textcolor{Plum}{- - -}). All other simulations parameters match Figure \ref{subfig:bdimlow}.
        (b) Comparing flux decline across simulations of \textit{B. diminuta} filtered through 200-nm pores at a base flux of $2.2 \times 10^{-4} \, \SI{}{\meter \per \second}$ (\textcolor{Blue}{$- \cdot - \cdot$}), at half the base flux (\textcolor{Dandelion}{$\cdot \cdot \cdot$}), and at double the base flux (\textcolor{NavyBlue}{$\cdot \cdot \cdot$}). All other parameters are the same as Figure \ref{subfig:bdimhigh}.}
        \label{fig:ffvars}
    \end{figure*}

In Figure \ref{fig:bdimlow-conc}, we run forward filtration simulations altering the feed water composition by changing the bacterial concentration. We compare the flux decline based on the experimental parameters shown in Figure \ref{subfig:bdimlow} to simulations in which the bacterial concentration is halved (\textcolor{Dandelion}{$\cdot \cdot \cdot$}) or doubled (\textcolor{NavyBlue}{$\cdot \cdot \cdot$}). Similar to the trends observed across the four experiments in Figure \ref{fig:ffsims}, the simulation with half the base concentration (\textcolor{Dandelion}{$\cdot \cdot \cdot$}) produces significantly less flux decline after the full filtration duration, while the simulation with double the base concentration (\textcolor{NavyBlue}{$\cdot \cdot \cdot$}) produces significantly more flux decline. Moreover, by keeping the bacterial concentration at the base level but swapping \textit{B.~diminuta} for \textit{S.~marcscens} (\textcolor{Plum}{- - -}), the flux experiences a rate of decline comparable to that observed when doubling the \textit{B.~diminuta} bacterial concentration. This makes sense when considering that \textit{S.~marcescens} has a cross-sectional area approximately 2.3 times larger than the cross-sectional area of \textit{B.~diminuta}, as shown in Figure \ref{fig:bacteria}.

In Figure \ref{fig:bdimhigh-flux}, we run forward filtration simulations altering the initial flux. We compare the normalized flux decline shown in Figure \ref{subfig:bdimhigh} to simulations in which the initial flux is halved (\textcolor{Dandelion}{$\cdot \cdot \cdot$}) or doubled (\textcolor{NavyBlue}{$\cdot \cdot \cdot$}). As the initial flux increases, so does the concavity of the flux decline curve, suggesting more rapid accumulation of foulants on the membrane. \textcolor{black}{Conversely, when halving the flux, initial concave down behavior appears due to the longer time needed for foulants to reach and begin accumulating on the membrane.}

Compared against the experimental data, these simulations demonstrate the flexibility and accuracy of our Monte Carlo model in capturing flux decline trends in forward filtration with different bacteria species, bacterial concentrations, membrane pore sizes, filtration durations, and applied transmembrane pressures.

\newpage
\subsection{Forward filtration-backwashing cycling}\label{subsec:FF-BW_validation}

We compare results for flux decline and recovery during forward filtration-backwashing (FF-BW) cycles produced by our Monte Carlo model to the constant-pressure, dead-end experiments performed by Gamage and Chellam with freshwater samples from Lake Houston \cite{article:LakeHoustonCycles:GamageChellam2014}. As with \cite{article:ConstantPressureDeadEnd:XuChellam2005}, the experiments were completed in commercial stirred cells, set up in their unstirred dead-end mode. In these experiments, raw (untreated) water and water pretreated with aluminum elecroflotation underwent five FF-BW cycles with length dictated by water production volume \cite{article:LakeHoustonCycles:GamageChellam2014}; that is, the researchers filtered 
$0.1 \,\SI{}{\liter}$ of water during every forward filtration phase using membranes with pore sizes $2.20 \times 10^{-9} \ \SI{}{\meter}$ \cite{article:LakeHoustonCycles:GamageChellam2012, article:LakeHoustonCycles:GamageChellam2014}. Due to foulant accumulation, the time to filter the same water quantity lengthened with each cycle. This is different from the forward filtration-only experiments, which filtered $0.1 \,\SI{}{\liter}$ of water only once.

According to empirical fits to constant-pressure blocking laws, the raw water experiments did eventually transition from intermediate blocking to cake filtration, while the pretreated water experiments were modeled solely by cake filtration \cite{article:LakeHoustonCycles:GamageChellam2012}. Since our model is built for intermediate blocking-dominant filtration scenarios, we focus our efforts on only the raw water experiments, imposing the intermediate blocking law throughout and ignoring for now the cake filtration transition, which would necessitate the development of a combined intermediate blocking-cake filtration model. Future work in this direction is discussed in Section \ref{sec:concl}.

We simulate five FF-BW cycles mimicking the experimental parameters and compare the resulting flux decline and recovery curves. The experimental data provides details on the duration of each FF phase but does not specify the duration of the BW phases. In our Monte Carlo simulations, we set the backwashing time in the first cycle to be 800 \SI{}{\second}, which is the time needed for $0.1 \,\SI{}{\liter}$ of water to flow through the clean experimental membrane area at the initial backwashing flux, $J_0^{\textnormal{BW}}$ (\SI{}{\meter \per \second}). For subsequent cycles, we incrementally increase the backwashing duration to account for the irreversibly attached foulant left on the membrane that slow down the process.

\begin{figure}[H]
    \centering
    \includegraphics[width=0.6\linewidth]{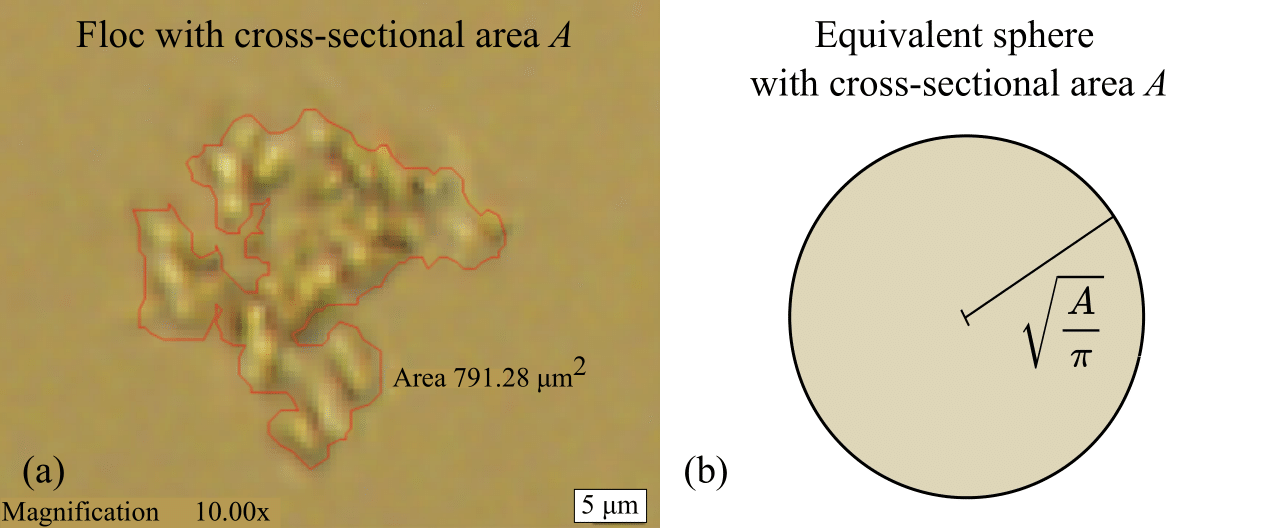}
    \caption{(a) Floc size analysis done using cellSens Dimension software on one model floc from the Lake Houston feed water. (b) Flocs are represented in the simulations of Figure \ref{fig:fluxcycle} as a sphere having equal cross-sectional area. Experiments filtering water from Lake Houston are reported in \cite{article:LakeHoustonCycles:GamageChellam2012, article:LakeHoustonCycles:GamageChellam2014}. The floc size analysis was described in more detail in \cite{article:LakeHoustonCycles:GamageChellam2012}, while the volume-based distribution of the flocs in the raw lake water was reported in more detail in \cite{article:LakeHoustonCycles:GamageChellam2014}. The floc image is from \cite{article:LakeHoustonCycles:GamageChellam2012}, edited for clarity and used with permission from Elsevier.}
    \label{fig:flocs}
\end{figure}

Since here the feed is a lake water sample and not custom-made to a desired bacterial concentration, as done for \textit{B.~diminuta} and \textit{S.~marcescens} in the experiments discussed in section \ref{subsec:FF_validation}, we expect multiple foulants, including bacteria, sediments, algae, and non-dissolved solids, to be contaminating the feed. While our model set-up can accommodate the introduction of multiple foulant types with different concentrations and sizes, data available on the raw lake water used does not include such measurements \cite{article:LakeHoustonCycles:GamageChellam2014}.
Gamage and Chellam report that their Lake Houston samples present foulants in flocs with volume-based mean diameters of $23.28 \times 10^{-6} \pm 3.18 \times 10^{-6}$ \SI{}{\meter} in the raw water. In reality, flocs exhibit a variety of non-uniform shapes; see Figure \ref{fig:flocs} for reference. A volume-based distribution of particle sizes re-interprets the non-uniform flocs as spheres having volumes equal to those of the original flocs, which are measured using image analysis software, light scattering, or some other technique \cite{web:VolumeDistribution:Horiba2010, web:VolumeDistribution:Ambivalue2017}. Based on this, we choose to model foulant particles as spheres in our FF-BW simulations and use $23.28 \times 10^{-6}$ \SI{}{\meter} as the diameter of the floc particles, as shown in Figure \ref{fig:flocs}.

\begin{figure}[H]
        \centering
          \includegraphics[width=0.75\linewidth]{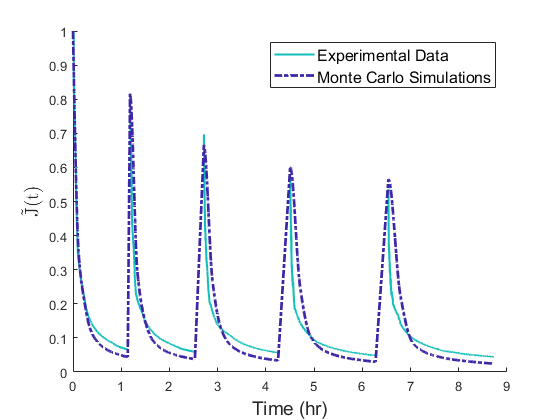}%
        
        \caption{Simulation of raw lake water microfiltration cycling (\textcolor{blue}{$- \cdot - \cdot$}) compared to experimental data from Gamage and Chellam (\textcolor{cyan}{---}) \cite{article:LakeHoustonCycles:GamageChellam2014}. Experimental water sample was sourced from the Lake Houston Canal at the City of Houston’s East Water Purification Plant in January 2011, and filtered through a membrane with 220-nm pores at a constant TMP of $14 \times 10^3$ \SI{}{\pascal} (2.03 psi) and initial flux of $2.53 \times 10^{-4} \SI{}{\meter \per \second}$. For the backwashing phases, the imposed TMP is $17 \times 10^3$ \SI{}{\pascal} (2.47 psi), corresponding to a flux of $3.06 \times 10^{-4}$ \SI{}{\meter \per \second} \cite{article:LakeHoustonCycles:GamageChellam2014}. The simulation curves are the average of 100 runs. Simulation parameters are detailed in Table \ref{tab:cycleparams}.}
        \label{fig:fluxcycle}
    \end{figure}

Even with the described approach to model flocs, it is not immediately apparent what the relevant $F_{\textnormal{conc}}$ would be for these Lake Houston samples.
The turbidity measurement for the raw water sample is provided as $14.7 \pm 1.5$ NTU \cite{article:LakeHoustonCycles:GamageChellam2014}. However, turbidity is an imperfect measure of the volumetric foulant concentration in contaminated water; while its scope somewhat intersects with that of total suspended solids (TSS), which more straightforwardly connects to $F_{\textnormal{conc}}$, a broadly applicable equation to relate the two does not exist, as the correlation between them varies by body of water, sampling location, and time \cite{web:TurbidityTSS:ThacktonPalermo2000}. Due to lack of data on the mass or quantity of suspended solids in the lake water and its correlation with the measured turbidity values, which would enable a rough calculation, we instead fit $F_{\textnormal{conc}}$ for Figure \ref{fig:fluxcycle} using the normalized residual introduced in equation (\ref{eqn:res}). 

Fitted values for $A_{\textnormal{adj}}$ and $F_{\textnormal{conc}}$ are needed to simulate the forward filtration phases in Figure \ref{fig:fluxcycle}. For the FF-BW cycling experiments, we also need to fit the backwashing parameter $p_{rem}$, not needed for the forward filtration simulations in section \ref{subsec:FF_validation}. Recall that $p_{rem}$ relays the maximum probability of removal when a fluid parcel encounters a grafted foulant. This parameter, which depends on foulant characteristics since some foulants are harder to remove than others, determines how well the flux recovers after a backwashing phase and, as discussed, $p_{rem} \ll p_{graft}$. We fit $A_{\textnormal{adj}}$ simultaneously with $F_{\textnormal{conc}}$ to the data from the FF phase of the first cycle; then, we fit $p_{rem}$ to the data from all of the FF-BW cycles. We fit $A_{\textnormal{adj}}$ and $F_{\textnormal{conc}}$ separately from $p_{rem}$ because the former two influence FF directly and BW indirectly while the latter influences BW directly and FF indirectly.

Table \ref{tab:cycleparams} lists the parameter values used to produce the simulations displayed in Figure \ref{fig:fluxcycle} (dashed), alongside the experimental results from Gamage and Chellam (solid) \cite{article:LakeHoustonCycles:GamageChellam2014}. As done earlier, we run 100 simulations at 0.1\% of the original experimental size and average the results.

We see that, even with our intermediate blocking-dominant simplification, the forward filtration branches in Figure \ref{fig:fluxcycle} qualitatively track those of the data well, and the backwashing curves return the flux to approximately the same level as is reached in the experiments. Since backwashing cannot recover all flux lost in the previous forward filtration phase due to irreversible attachments and remaining fouling \cite{article:RemainingFouling:RemizeETAL2010, article:IrreversibleFouling:GuptaChellam2022}, we expect the overall flux to trend downward, as observed in the experimental and simulation data in Figure \ref{fig:fluxcycle}.

Altogether, these results comparing numerical simulations to data from forward filtration-only and FF-BW cycle experiments show that our model can faithfully capture single- and multi-cycle flux decline and recovery behaviors in constant-pressure, dead-end filtration operations where intermediate blocking is the dominant fouling mechanism. We accomplish this through Monte Carlo simulations built on intermediate pore blocking laws and real-world parameters. The strong connection to experimentally measurable quantities and the good fit of our simulation results suggests the viability of our model for predicting the flux patterns under different filtration conditions.

\begin{table}[t]
\caption{\label{tab:cycleparams}%
In-model parameters and variables used for simulations in Figure \ref{fig:fluxcycle}. Most of the physical parameter values were derived or inferred from the original experiments \cite{article:LakeHoustonCycles:GamageChellam2012,article:LakeHoustonCycles:GamageChellam2014} or numerical experiments done by the same research group \cite{article:OptimalControlBackwash:CoganChellam2014}.}
\centering
\begin{tabular}{llll}
\hline
  & \textrm{Description (units)} & 
\textrm{Fig. \ref{fig:fluxcycle}} & Source(s) \\
\hline
    $J_0$ & Initial flux (\SI{}{\meter \per \second}) & $2.53 \times 10^{-4}$ & \cite{article:LakeHoustonCycles:GamageChellam2014} \\
    $J_0^{\textnormal{BW}}$ & Backwashing flux (\SI{}{\meter \per \second}) & $3.06 \times 10^{-4}$ & \cite{article:LakeHoustonCycles:GamageChellam2014} \\

    $\Pen$ & P{\'e}clet number & 15812.5 & - \\
    $F_{\textnormal{conc}}$ & Foulant concentration (cells \SI{}{\per \meter \cubed}) & $1.41 \times 10^{11}$ & Model fit (with $A_{\textnormal{adj}}$) \\
    $F_{\textnormal{area}}$ & Cross-sectional foulant area (\SI{}{\meter \squared} cell$^{-1}$) & $4.26 \times 10^{-10}$ & \cite{article:LakeHoustonCycles:GamageChellam2012,article:LakeHoustonCycles:GamageChellam2014} \\
    $A_{\textnormal{adj}}$ & Model parameter (-) & 1.82 & Model fit (with $F_{\textnormal{conc}}$) \\
    $t_F$ & Forward filtration duration (\SI{}{\second}) & \textit{varies} & \cite{article:OptimalControlBackwash:CoganChellam2014} \\
    $t_{BW}$ & Backwashing duration (\SI{}{\second}) & \textit{varies} & \cite{article:OptimalControlBackwash:CoganChellam2014} \\
    $p_{rem}$ & Maximum probability of removal (-) & 0.0071 & Model fit \\
    $d_{\textnormal{pore}}$ & Pore diameter (\SI{}{\meter}) & $2.2 \times 10^{-7}$ & \cite{article:LakeHoustonCycles:GamageChellam2014} \\
    n & Input rate (cells~\SI{}{\per \second}) & 15 & - \\
    \hline
\end{tabular}
\end{table}

Our simulations rely primarily on integration of real-world parameters in the model. For instance, the in-channel foulant starting count $F_0$ and the input number $n$ at each time step depend on knowledge of the experimental feed foulant concentration $F_{\textnormal{conc}}$, and the flux equation depends on knowledge of the shape and average dimensions of the foulants. We believe that the dependence on physical parameters rather than estimated model parameters is a strength of the model because it facilitates straightforward modification of inputs to simulate many other real-world filtration scenarios. However, we are cognizant that detailed experimental information may not always be available, as was the case in the cycling experiments. As such, here we have shown how to handle two very different datasets---one in which all desired physical parameters were known and one in which key physical parameters were missing---by implementing parameter fitting techniques for multiple parameters simultaneously, thus enhancing the flexibility of our model.

\section{\label{sec:concl}Conclusions}

In this paper, we developed a Monte Carlo model that relied on physical parameters to simulate forward filtration and backwashing through dead-end, flat-sheet membranes in constant-pressure filtration operations. Based on the timescales and foulant compositions employed, we assumed that intermediate blocking was the dominant fouling mechanism at play, and derived a model equation for the flux decline and recovery that results from membrane fouling and backwashing.

We validated the forward filtration (FF) phase of the model against microfiltration experiments conducted with the bacteria species \textit{B.~diminuta} and \textit{S.~marcescens} \cite{article:ConstantPressureDeadEnd:XuChellam2005} and validated forward filtration-backwashing (FF-BW) cycling against filtration experiments conducted with lake water samples \cite{article:LakeHoustonCycles:GamageChellam2014}. We observed good agreement between the experimental data and the flux behavior captured by our numerical simulations.

Our model can be improved in future work in a few notable ways. First, our expression for flux decline, equation (\ref{eqn:sim_intermediate_flux}), describes the intermediate blocking law of constant-pressure filtration; this is the filtration period during which there remains membrane surface area open for direct attachment of the bacteria. Once foulants have no opportunity to graft directly onto the membrane due to previous foulant accumulation, we theoretically enter the cake filtration phase, for which a different flux expression is needed \cite{article:MembraneBlockingFluxDecline:BowenCalvoHernandez1995, article:ReviewPoreBlocking:Iritani2013}. A natural next step is to develop a flux decline equation to describe membrane fouling that includes cake formation. A classical cake blocking model exists \cite{article:ConstantPressureFiltration:Hermia1982}, as well as more recent theoretical expressions that bridge the intermediate blocking law and cake filtration \cite{article:BlockageCake:HoZydney2000, article:CombinedFouling:BoltonETAL2006, article:CombinedFouling:DuclosOrselloETAL2006, article:CombinedFouling:HouETAL2017, article:CombinedFouling:SanaeiCummings2019}. In the FF-only scenarios examined in this work, this transition does not occur; and in the FF-BW cycles with raw lake water, we ignored cake filtration. The build-up of a bacterial cake on filters is a prominent problem in membrane filtration, and thus worth integrating into the model; this addition would enable us to more accurately represent the fouling regimes of Figure \ref{fig:fluxcycle} and investigate cake filtration-dominated experiments (e.g., \cite{article:LakeHoustonCycles:GamageChellam2012, article:LakeHoustonCycles:GamageChellam2014}).

The decision to use a 2D model domain presents a limitation to what we are able to visualize with the simulations. Since the interest of the current study was in capturing flux decline and recovery due to foulant accumulation and removal, it was prudent and efficient to simplify our domain to 2D. However, extending our model to 3D would allow for the visualization of foulant build-up morphology and patterns formed on the membrane.

Lastly, the lack of foulant concentration information discussed in section \ref{subsec:FF-BW_validation} is very common in most real-world cases due to the difficulty and cost of collecting it. We relied on least squares fitting to approximate the foulant concentrations for Figure \ref{fig:fluxcycle}, as well as $A_{\textnormal{adj}}$ for all simulations. Future work could improve the optimization routines applied to estimate these and other parameters, as the available data requires.

Through this work, we have demonstrated the applicability of our Monte Carlo method-based model for the simulation of foulant transport, attachment, and removal in dead-end, constant-pressure membrane filtration experiments conducted in the context of biofouling studies \cite{article:ConstantPressureDeadEnd:XuChellam2005} and surface water treatment \cite{article:LakeHoustonCycles:GamageChellam2014}. As the principles of membrane fouling and regeneration are similar in other industries, including food manufacturing and the pharmaceuticaul industry, we could easily translate the model for application in other contexts.

\section*{Acknowledgements}
We thank Shankararaman Chellam at Texas A\&M University and Nicholas G.~Cogan at Florida State University for providing original experimental data and feedback on early versions of the model.
This research was supported by NSF CBET-2211001.

\vspace{1cm}

{\bf CRediT Author Statement.} A.R.D.: Conceptualization, Data Curation, Formal Analysis, Investigation, Methodology, Software, Validation, Visualization, Writing--Original Draft, Writing--Review \& Editing.
F.B.: Conceptualization, Formal Analysis, Funding Acquisition, Investigation, Methodology, Project Administration, Resources, Supervision, Writing--Review \& Editing.

\bibliographystyle{abbrv}
\bibliography{bibfile}

\end{document}